\documentclass{optica-article}

\journal{opticajournal} % for journals or Optica Open

\articletype{Research Article}

\usepackage{lineno}
\usepackage{amsmath}
\usepackage{soul}

\usepackage{graphicx}% Include figure files
\usepackage{dcolumn}% Align table columns on decimal point
\usepackage{bm}% bold math
\usepackage{amsmath}
\usepackage{braket}
\usepackage{mathtools}
\usepackage{float}

\usepackage{xcolor}
\newcommand{\BS}[1]{\textcolor{black}{ #1}}

\newcommand{\WB}[1]{\textcolor{black}{ #1}} %Wagner
\DeclareUnicodeCharacter{0301}{\'{e}}
\begin{document}

\title{Light correcting light with nonlinear optics} %Very preliminary title

\author{Sachleen Singh\authormark{1$\dagger$}, Bereneice Sephton\authormark{2$\dagger$*}, Wagner Tavares Buono\authormark{1}, Vincenzo D'Ambrosio\authormark{2}, Thomas Konrad\authormark{3} and Andrew Forbes\authormark{1}}

\address{\authormark{1}University of the Witwatersrand, Johannesburg, South Africa\\
\authormark{2}University of Naples Federico II, Naples, Italy \\
\authormark{3}University of KwaZulu-Natal, Durban, South Africa \\
\authormark{$\dagger$}Equal contribution to the work}

\email{\authormark{*}bereneice21@gmail.com} %% email address is required; see note below about the corresponding author designation

% use {asbstract*} to suppress the copyright line. Copyright information will be added in production

\begin{abstract*} 
Structured light, where complex optical fields are tailored in all their degrees of freedom, has become highly topical of late, advanced by a sophisticated toolkit comprising both linear and nonlinear optics.  Removing undesired structure from light is far less developed, leveraging mostly on inverting the distortion, e.g., with adaptive optics or the inverse transmission matrix of a complex channel, both requiring that the distortion is fully characterised through appropriate measurement.  Here we show that distortions in spatially structured light can be corrected through difference frequency generation in a nonlinear crystal without any need for the distortion to be known.  We demonstrate the versatility of our approach by using a wide range of aberrations and structured light modes, including higher-order orbital angular momentum (OAM) beams, showing excellent recovery of the original undistorted field.  To highlight the efficacy of this process, we deploy the system in a prepare-and-measure communications link with OAM, showing minimal crosstalk even when the transmission channel is highly aberrated, and outline how the approach could be extended to alternative experimental modalities and nonlinear processes.  Our demonstration of light correcting light without the need for measurement opens a new approach to measurement-free error correction for classical and quantum structured light, with direct applications in imaging, sensing and communication.

\end{abstract*}

%%%%%%%%%%%%%%%%%%%%%%%%%%  body  %%%%%%%%%%%%%%%%%%%%%%%%%%
\section{Introduction}
Light, and with it, the transverse tailoring of phase and amplitude to create the so-called structured light \cite{forbes2021structured}, presents a large field of active research with wide ranging applications \cite{he2022towards}, from optical trapping \cite{yang2021optical} to communication \cite{willner2021orbital}. The toolkit has become highly versatile covering generation, control and detection schemes that include liquid crystals \cite{lazarev2019beyond,rubano2019q}, digital micromirror devices \cite{scholes2020structured} and metasurfaces \cite{dorrah2022tunable}.  Beyond linear optics, structured light control with nonlinear optics has become topical of late \cite{buono2022nonlinear}, shifting the focus of attention from wavelength change and efficiency to spatial modal creation, control and detection.  This has led to a re-invention of the field with a modern twist, ushering in new selection rules \cite{wu2023observation,chen2020phase,tang2020harmonic} and processes \cite{wu2022conformal,luttmann2023nonlinear,da2022observation,da2021stimulated} while fostering wide reaching applications, including spatial mode creation \cite{hancock2019free,steinlechner2016frequency,zhou2014generation} and detection \cite{sephton2019spatial,xu2023orthogonal}, image processing \cite{schlickriede2020nonlinear,qiu2018spiral,ribeiro2001observation,paterova2020hyperspectral} and filtering \cite{rocha2021speckle}, holography \cite{trajtenberg2015axis,almeida2016nonlinear,fang2021high}, enhanced interferometry \cite{zhang2023real}, high-dimensional teleportation \cite{sephton2021stimulated, qiu2021quantum}, as well as the development of modern nonlinear materials \cite{wei2018experimental,xu2022femtosecond,zhang2021nonlinear,guo2023ultrathin}. 

Unfortunately the spatial structure of light becomes distorted in complex channels \cite{lib2022quantum,gigan2022roadmap,rotter2017light,cao2022shaping}, arresting its full potential.  Although phase conjugation of structured light is possible by nonlinear optics \cite{de2019real}, it does not correct the distortion but rather produces the negative of it, requiring a time reversal step \cite{fisher2012optical}.  To mitigate these drawbacks, a measurement based approach to structured light correction is now ubiquitous, for example, using adaptive optics \cite{sorelli2019entanglement,grunwald2022high,booth2014adaptive,dai2019active,hampson2021adaptive} and wavefront shaping \cite{cheng2023high}, inversion of the transmission matrix of complex channels \cite{bachmann2023highly,popoff2010measuring,vellekoop2008universal}, and finding invariances that remain distortion-free \cite{klug2023robust,nape2022revealing,pai2021scattering}. 

Here we show that light can correct light without the need for any measurement.  We exploit parametric wave mixing by difference frequency generation in a nonlinear crystal to restore the information encoded into the structure of light, even after it has passed through a highly aberrating channel. In order to achieve this, two input beams, one with information encoded into its structure and the other as a probe, are passed through the same aberrating channel followed by difference frequency generation  in a nonlinear crystal, returning only the desired information. This is due to the nature of the parametric wave mixing process which outputs the product of one of the input modes with the conjugate of the other.
We demonstrate the versatility of our approach by using a wide range of aberrations and structured light modes, from Gaussian beams to orbital angular momentum (OAM) beams and their superpositions, showing excellent recovery of the original undistorted field.  To highlight the efficacy of this, we consider the crosstalk matrix of a 15 dimensional OAM alphabet across a noisy channel comprising an arbitrary aberration, showing very good recovery of the information.  We outline how our approach can be used across multiple wavelengths that could be close or far apart, offering a new approach to measurement-free error correction for classical and quantum structured light.

\section{Concept}

\begin{figure}[t!]
\centering\includegraphics[width=8cm]{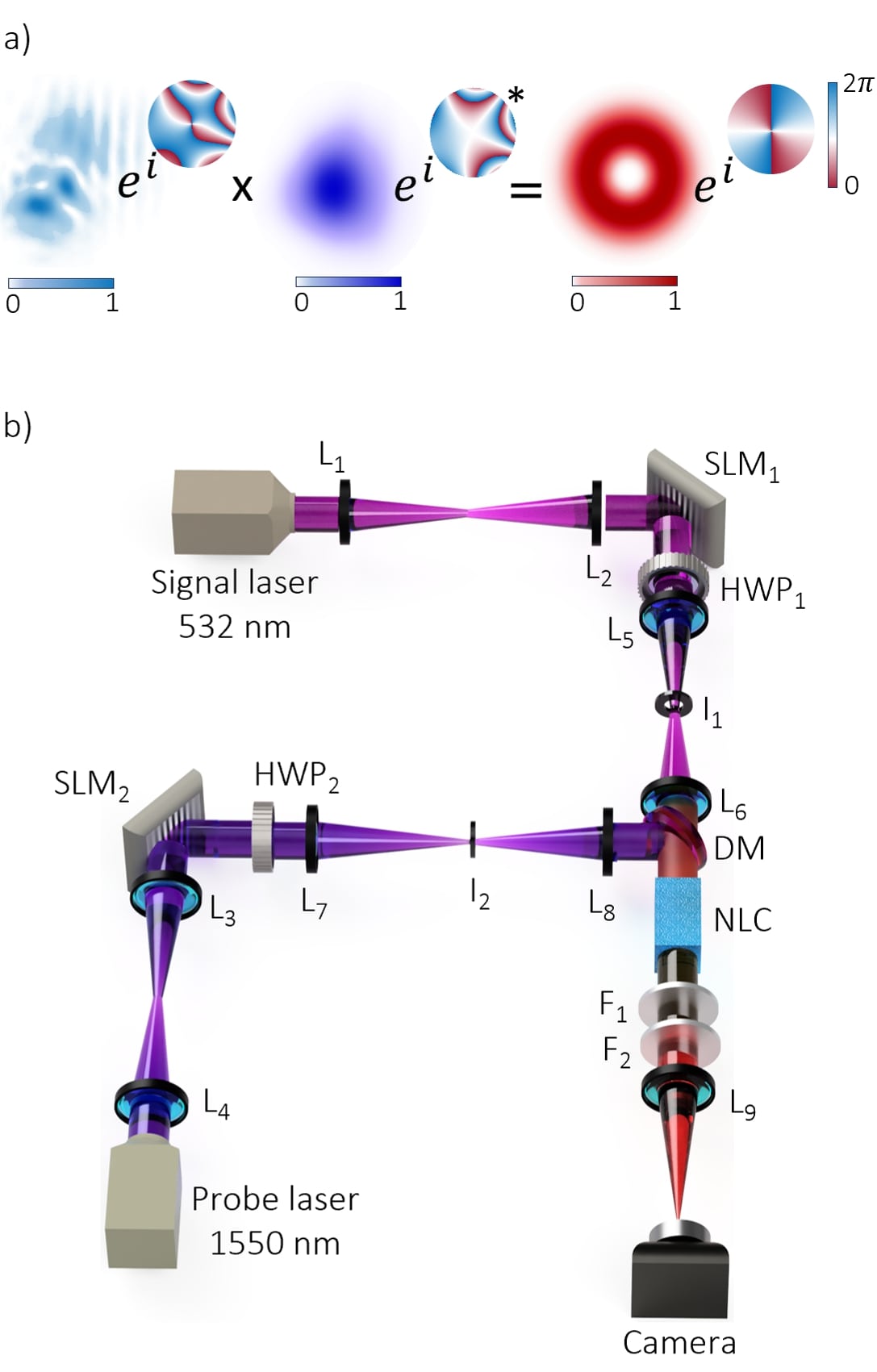}
\caption{(a) Concept of correcting aberrated states by using light to correct light. The product of an input beam (middle mode) with another containing the same phase aberration (exponential term) cancels the identical distortion present in the structure carried by a second input beam (left mode) to restore the unaberrated state (right mode) in the difference-frequency beam generated from nonlinear wave mixing. Beams are shown in the far-field for conceptual clarity. (b) Experimental setup used to apply and correct distortions on structured modes with difference frequency generation. SLM, spatial light modulator; HWP, half-wave plates; I, apertures; DM, dichroic mirror; NLC, nonlinear crystal; F$_1$, shortpass and F$_2$ longpass wavelength filters; L$_1$ (18 mm), L$_2$ (200 mm), L$_3$ (300 mm), L$_4$ (75 mm), L$_5$ (500 mm), L$_6$ (100 mm), L$_7$ (750 mm) and L$_8$ (100 mm) are lenses. }
\label{Fig1}
\end{figure}

With difference frequency generation (DFG), two electric fields ($\textbf{E}_1$ and $\textbf{E}_2$) mix in a second order nonlinear crystal to generate a third beam ($\textbf{E}_{\text{G}}$). Here, each field possesses a transverse spatial structure $M_n(r,\phi)$ and polarisation, indicated by the unit vector $\hat{\boldsymbol{e}}_n$,
\begin{equation}
    \textbf{E}_{n} = M_n(r,\phi)\hat{\boldsymbol{e}}_n
\end{equation}
where $n = \{1,2,\text{G}\}$ refers to the first, second and generated beams, while ($r,\phi$) are the radial and azimuthal coordinates in the transverse spatial plane. Coherent amplification of the generated field occurs along the crystal length when the phase-matching conditions are satisfied. This applies a constraint between the wave-vectors and interacting fields \cite{zhang2017phase}, ensuring conservation of energy and momentum in the process. For DFG, the energy of the generated field in the paraxial regime is aptly given by the difference of the input angular frequencies, $\omega_1 - \omega_2 = \omega_{\text{G}}$ and wavevectors, $\boldsymbol{k}_1 - \boldsymbol{k}_2 = \boldsymbol{k}_\text{G}$ for the transverse components of the interacting fields. A sufficiently large bandwidth for phase-matching of the longitudinal component in the thin-crystal limit causes the spatial profile of the generated field to be reduced to the product of the two input fields \cite{bloembergen1980conservation}. Following from the conservation rules, the output field then holds the combined information of the input fields such that the spatial structure of the generated field is proportional to that of the first input and the complex conjugate of the second input,
\begin{equation}
        M_{\text{G}} = \eta M_{1}M^{*}_{2},
\end{equation}
where $\eta$ is a constant related to the efficiency of the process and $*$ indicates complex conjugation.

By considering the complex form of the spatial structures at the beam waist (neglecting propagation terms for simplicity), $M_{n} = A_{n}(r,\phi)e^{i\Phi_{n}(r,\phi)}$ where $A(r,\phi)$ is the amplitude and $\Phi(r,\phi)$ the phase, the effect of DFG is to conjugate the phase distribution of the second beam and adding it to the phase distribution of the first, $\Phi_{\text{G}} = \Phi_{1} + (-\Phi_{2})$. Where the second beam's phase is uniform or null, the generated beam phase is simply that which is carried by the first beam ($\Phi_{\text{G}} = \Phi_{1}$). As a result, the generated beam will contain any desired structure ($\Phi_\text{info}$), that the first beam contains. This, however, is true for any additional distortions ($\Phi_\text{Ab}$) experienced by the beam as well. For such an event, the generated beam will then have a phase of $\Phi_\text{G} = \Phi_\text{info}+\Phi_\text{Ab}$, such that the modal information or purity is degraded and seen in distortion of the intensity profile upon propagation. One may now consider the case where the contribution of the second beam can be exploited.  Without loss of generality, we consider an example where we seek to restore Laguerre-Gaussian (LG$_{\ell}$) modes of zero radial index ($p = 0$) and abitrary $\ell$, from an aberrated state. Figure \ref{Fig1} (a) illustrates this concept. Notably, these structured modes hold orbital angular momentum (OAM) as a degree of freedom and are characterised by the integer parameter $\ell$, which yields $\ell \hbar$ OAM per photon and $\ell$ number of twists in the phase-front per wavelength (red to blue transitions in the rightmost phase inset). To correct the aberration, one need only see that by using the same aberration phase on the second beam, the original helical phase can be restored. The distortions of the LG mode amplitude (depicted alongside the phase terms) is then corrected to reveal the characteristic doughnut intensity distribution. Here, due to the naturally occurring phase conjugation in the crystal, the initial disturbance, e.g., $M_1 = \text{LG}_{(\ell = 2)}e^{i\Phi_\text{Ab}}$, also present in the second beam (using a Gaussian profile to conserve the structure of the first beam), e.g., $M_2 = \text{LG}_{(\ell = 0)}e^{i\Phi_\text{Ab}}$, cancels the distortion in the generated beam, $\Phi_{\text{G}} = (\Phi_\text{info} + \Phi_\text{Ab}) - \Phi_\text{Ab} =  \Phi_\text{info}$, while preserving the initial phase and amplitude. Cancellation of the unwanted disturbance, such as turbulence, and successful transfer of the desired structure carried by the first beam is therefore achieved by using the structure of one light beam to correct that of the other. Note that while we have outlined the concept in the near-field of the aberration, far-field scenarios of phase and amplitude coupling can simply be reversed by a lens before entering the crystal.

\section{Experimental Results}
To demonstrate the principle of using a second beam in DFG to correct phase aberrations on an initial input beam, we implemented the experimental setup as shown in Fig. \ref{Fig1} (b). Here, two continuous wave lasers of wavelengths 532 nm (VIS) and 1550 nm (IR) were collimated and expanded onto liquid crystal spatial light modulators (SLM$_1$, SLM$_2$), before demagnification and imaging onto a type-0 nonlinear crystal (NLC, periodically-poled KTP) with a 4$f$-lens system (L$_5$, L$_6$ and L$_7$, L$_8$). Complex amplitude modulation \cite{clark2016comparison} was used to encode the desired states of each input beam, which we will refer to as probe and signal beams to clarify their roles in the correction process. Apertures (I$_1$, I$_2$) in the Fourier plane spatially filtered the 1st order modulated light from the SLMs, respectively forming the \BS{signal} and \BS{probe} input modes. Half-waveplates (HWP$_1$, HWP$_2$) in each arm then respectively adjusted the polarisation for phase-matching and a dichroic mirror (DM) used to collinearly combine the beams before the NLC. A long- and short-pass wavelength filter (F$_1$, F$_2$) placed after the crystal isolates the DFG beam. The generated beam was then focused onto a camera by a lens in a 2$f$ configuration (F$_9$), detecting the Fourier plane of the DFG modes.

\begin{figure}[ht!]
\centering\includegraphics[width=8cm]{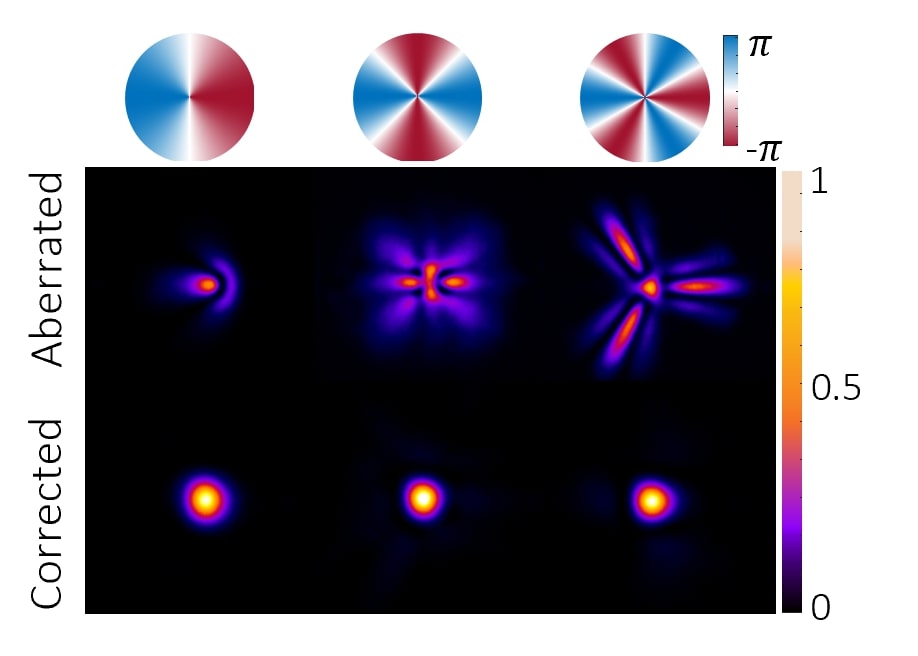}
\caption{Three azimuthal aberrations (top insets) were applied to a Gaussian signal beam, resulting in measured intensity distortions in the far-field (Aberrated row). Application of the nonlinear correction process with a probe beam results in the recovery of the initial Gaussian beam, as evident in the measured far-field intensities in the bottom row (Corrected). All intensities are normalized to 1.}
\label{Fig2}
\end{figure}

We now experimentally realise this concept with the results shown in Fig.~\ref{Fig2}. Here, three azimuthally-varying phase aberrations, $\Phi_\text{Ab} = \exp(i\pi\cos{(n\phi}))$ where $n = \{1,2,3\}$ (shown in the top insets) were applied to the IR Gaussian signal beam. \BS{The Gaussian structure and flat phase of the IR probe beam is retained for the process.} As expected, aberrations on the generated mode distorts the beams in the far-field, as seen in the top row. By employing the light correcting light approach with DFG, implemented by now applying the same aberrational phase to the VIS probe beam, we find the initial structure is corrected and confirmed with unaberrated Gaussian distributions in the bottom row.

\begin{figure}[htbp!]
\centering\includegraphics[width=16cm]{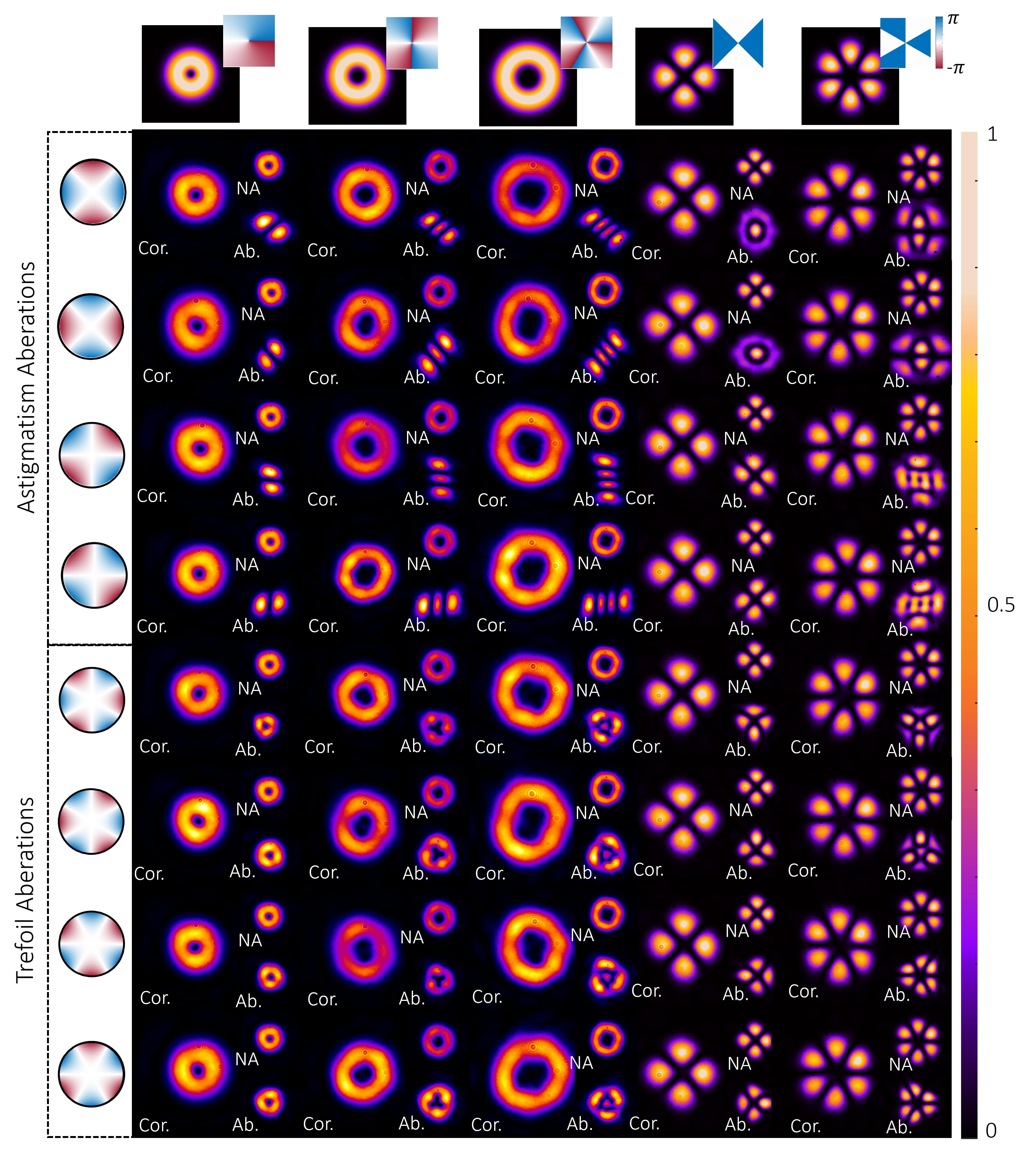}
\caption{Experimental correction of astigmatism and trefoil aberrations for five different spatial states (column-wise) where $\ell$ = \{1,2,3\} and petal modes with LG superpositions ($\frac{1}{\sqrt{2}}\left(\text{LG}_{\ell} + \text{LG}_{-\ell}\right)$) where $\ell$ = \{2,3\}. The correction has been applied for both vertical and oblique combinations of aberrations. Further every such combination has been corrected for both positive and negative strength coefficients. The applied phase distortion has been shown in left panel. Every experimental picture shows results for corrected (Cor.) mode with corresponding aberrated (Ab.) and not aberrated (NA) modes as insets. The expected simulated intensity and phase profiles has been shown in the top row.}
\label{Fig3}
\end{figure}
In Fig.~\ref{Fig3} we next explore aberrations having both radial and azimuthal dependence, while also expanding the encoded states to higher-order modes. We note any spatial modes may be used and chose LG due to their extensive applications from communications to metrology \cite{yao2011orbital,shen2019optical}. The Zernike basis (Z$_{m,n}$) \cite{lakshminarayanan2011zernike} with azimuthal frequency, $n$, and radial order, $m$, are used to simulate the unwanted distortions, forming a natural basis for optical aberrations \cite{love1997wave,wang1980wave}. Modes representing astigmatism ($Z_{2,2}, Z_{2,-2}$) and trefoil ($Z_{3,3}, Z_{3,-3}$) were then chosen from the Zernike family and applied with same aberrational strength. The expected doughnut intensity distributions of these LG states (first three panels), show good agreement to the unaberrated DFG intensities (NA, top-right of each modal set). After the structured signal beam encounters each aberration, however, significant deviations in the DFG intensity profiles (Ab.) are observed, obscuring the modes and related information. Applying the same phase distortion to the Gaussian probe shows successful restoration of the modal structure in the DFG beam by cancellation of the aberrational phase (Cor.). Applicability to states with more modal complexity is further demonstrated by constructing modes from a superpositon of LG states ($\frac{1}{\sqrt{2}}\left(\text{LG}_{\ell} + \text{LG}_{-\ell}\right)$), giving 0 to $\pi$ wedge phase steps with petal intensity structures. This is shown in the last two panels where $\ell$ = \{2,3\}, respectively. Similarly, aberrations caused notable distortions in the detected intensity distributions, but excellent restoration when applying our correction approach.

%%%%%%%%%%%%%%%%%%%%%%%%%%%%%%%%%%%%%%%%%%%%%%%%%%%%%%%%%%%%%%%%%%%%%%%%%%%%%%%%%%%%%%%%%%%%%%%%%%%%%%%%%%%%%%%
%%%%%%%%%%%%%%%%%%%%%%%%%%%%%%%%%%%%%%%%%%%%%%%%%%%%%%%%%%%%%%%%%%%%%%%%%%%%%%%%%%%%%%%%%%%%%%%%%%%%%%%%%%%%%%%
\begin{figure}[t!]
\centering\includegraphics[width=14cm]{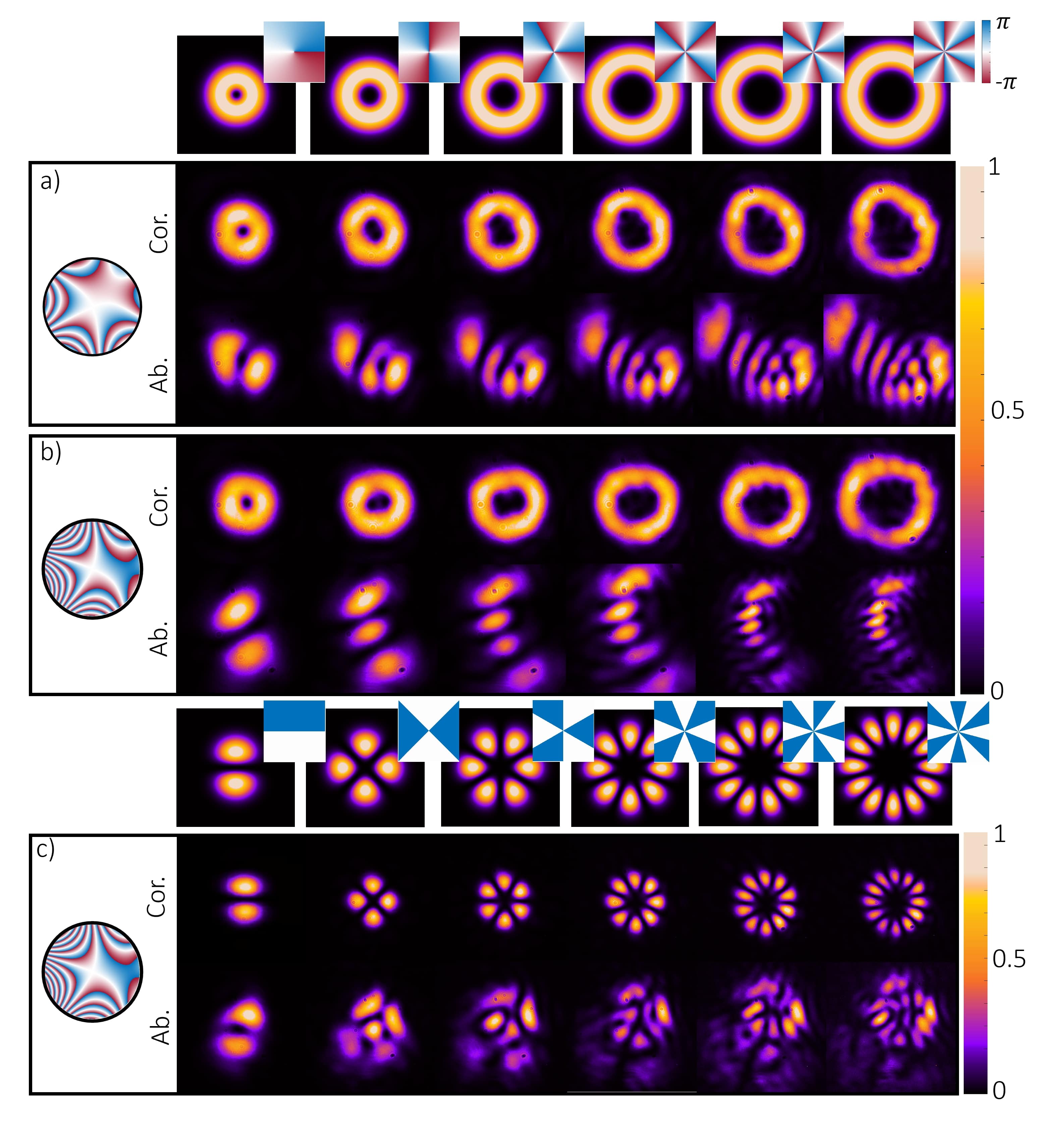}
\caption{Correction for superpositions of astigmatism and trefoil with arbitrarily chosen strengths. Left column insets show the aberrating phases acting on spatial modes (top insets). Experimentally aberrated (Ab.) and corrected (Cor.) far-field intensities for LG beams increasing colunmwise in OAM from $\ell = [1,6]$ are given for aberrations with (a) three and (b) four mode superpositions. (c) Experimental results with the same OAM range for the LG superpositions ($\frac{1}{\sqrt{2}}\left(\text{LG}_{\ell} + \text{LG}_{-\ell}\right)$) are given for the same four mode aberration.}
\label{Fig4}
\end{figure}
%%%%%%%%%%%%%%%%%%%%%%%%%%%%%%%%%%%%%%%%%%%%%%%%%%%%%%%%%%%%%%%%%%%%%%%%%%%%%%%%%%%%%%%%%%%%%%%%%%%%%%%%%%%%%%%
%%%%%%%%%%%%%%%%%%%%%%%%%%%%%%%%%%%%%%%%%%%%%%%%%%%%%%%%%%%%%%%%%%%%%%%%%%%%%%%%%%%%%%%%%%%%%%%%%%%%%%%%%%%%%%%

Greater aberrational complexity is also introduced by taking three, $\Phi_\text{Ab} = 5Z_{2,2} + 5Z_{2,-2} + 10Z_{3,3}$, and four, $\Phi_\text{Ab} = 10Z_{2,2} - 10Z_{2,-2} - 10Z_{3,3} + 10Z_{3,-3}$, mode superpositions of the Zernike basis states. This is shown in Fig.~\ref{Fig4} (a) and (b), respectively, where the signal beam was also encoded with LG modes of $\ell = \left[1:6\right]$. Here, deleterious distortions obscure the encoded doughnuts (top insets) into intermittent linear structures (bottom rows, Ab.). With the same phase distortion on the probe beam, we again find the output structure returns to the ring profile. While the modes are excellently restored, a reduction in the correction efficacy appears as the $\ell$ value increases. This can be attributed to an increase in the generated beam size of $w_\ell = w_0 \sqrt{(|\ell| +1)}$ where $w_{\ell}$ is the OAM beam waist and $w_0$ the waist of the fundamental Gaussian mode. As a result of increasing size, greater interaction with the optical elements occur, leading to the modes obtaining additional peripheral aberrations not \BS{encoded and} accounted for in the probe profile. % Despite the aberrational effect of the system, good agreement between the corrected and encoded profiles is still possible. 
In Fig.~\ref{Fig4} (c), the same four-mode aberration is applied to the previous petal superpositions where $\ell \in [1,6]$. The aberrating phases on the signal beam similarly destroy the DFG structure, such that they can no longer be identified in comparison to the expected distributions (top insets in (c)). Excellent agreement then occurs when the probe is used to correct for the distortion. While a small decrease in correction efficacy is also seen as $\ell$ increases, favourable restoration is still seen up to the largest state.
%%%%%%%%%%%%%%%%%%%%%%%%%%%%%%%%%%%%%%%%%%%%%%%%%%%%%%%%%%%%%%%%%%%%%%%%%%%%%%%%%%%%%%%%%%%%%%%%%%%%%%%%%%%%%%%
\begin{figure}[ht!]
\centering\includegraphics[width=8cm]{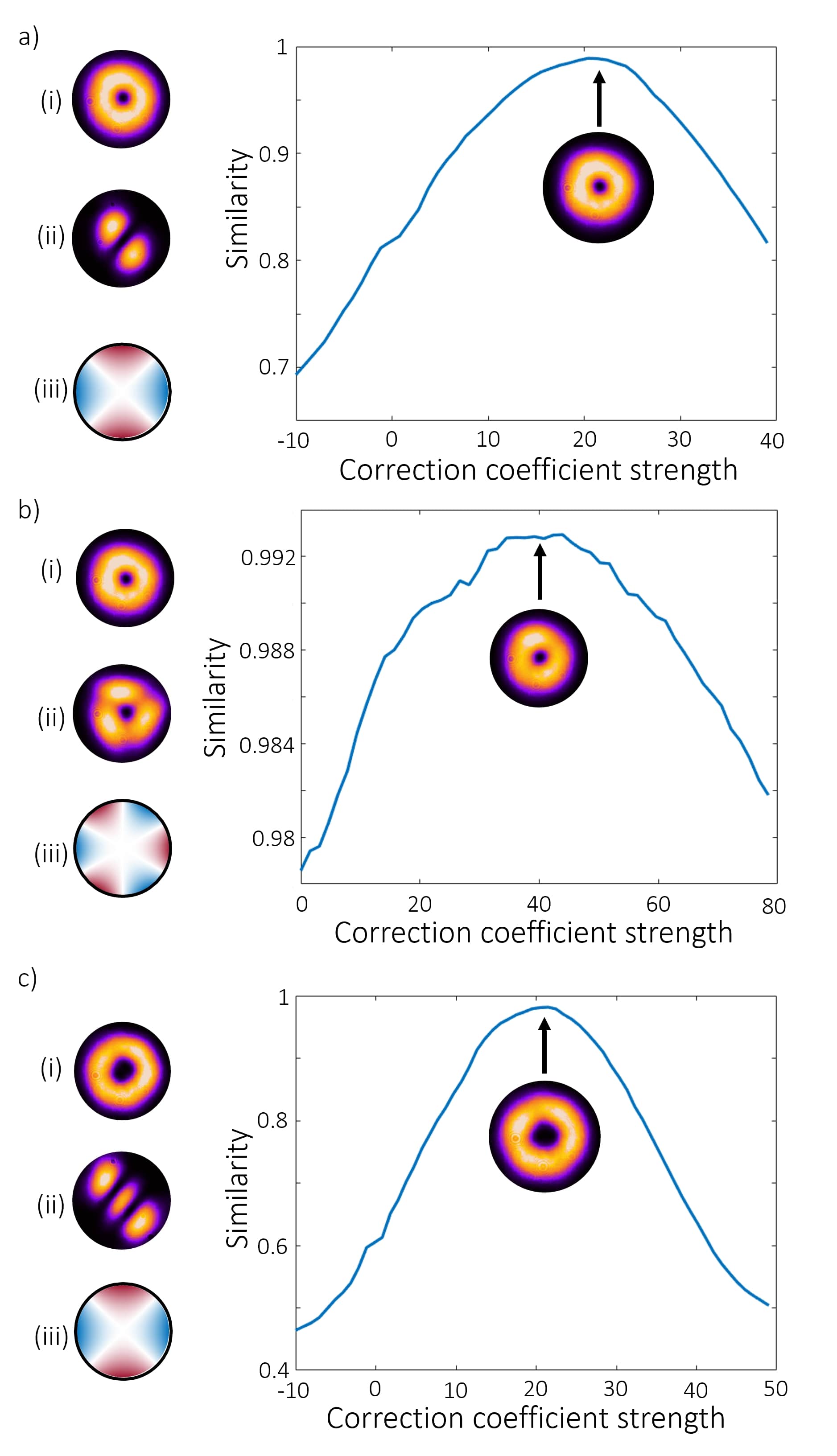}
\caption{Resizing beam changes the relative strength of the aberrations for (a) $\ell = 1$ with astigmatism (order 2) and coefficient of 10 (calculated mismatch = 1.443), (b) $\ell = 1$ with trefoil (order 3) and coefficient of 15 (calculated mismatch = 1.431), (c) $\ell = 2$ with astigmatism (order 2) and coefficient of 10 (calculated mismatch = 1.466). Rightmost insets show the (i) unberrated downconverted mode, (ii) aberrated downconverted mode (no correction) and (iii) aberrating Zernike mode phase distribution.}
\label{Fig5}
\end{figure}
%%%%%%%%%%%%%%%%%%%%%%%%%%%%%%%%%%%%%%%%%%%%%%%%%%%%%%%%%%%%%%%%%%%%%%%%%%%%%%%%%%%%%%%%%%%%%%%%%%%%%%%%%%%%%%%

In the practical application of our measurement-free approach, a necessary condition is both probe and signal beams incur identical phase distortions for proper cancellation. One may thus consider the wavelength-dependence of the phase accumulated by two beams of differing wavelengths (e.g. $\lambda_1$ and $\lambda_2$) traversing a distorting medium with refractive index $n_{\lambda}$ such as glass with thickness, $d$, varying across the beam profile. An unwanted dynamic phase of $\varphi = \frac{2\pi n_{\lambda}d}{\lambda}$ is subsequently imparted at each point across the spatial profiles such that the phases for one wavelength can then be related as $\varphi_{\lambda_{2}} = \alpha \varphi_{\lambda_{1}}$ to the other, where $\alpha = \frac{\lambda_{1}n_{\lambda_{2}}}{\lambda_{2}n_{\lambda_{1}}}$. The disparity is reduced to a constant value, fixed by the chosen wavelengths and traversed medium. Accordingly, one need only account for a difference in the strength for the same aberrational distribution.
%%%%%%%%%%%%%%%%%%%%%%%%%%%%%%%%%%%%%%%%%%%%%%%%%%%%%%%%%%%%%%%%%%%%%%%%%%%%%%%%%%%%%%%%%%%%%%%%%%%%%%%%%%%%%%%
\begin{figure}[b]
\centering\includegraphics[width=8cm]{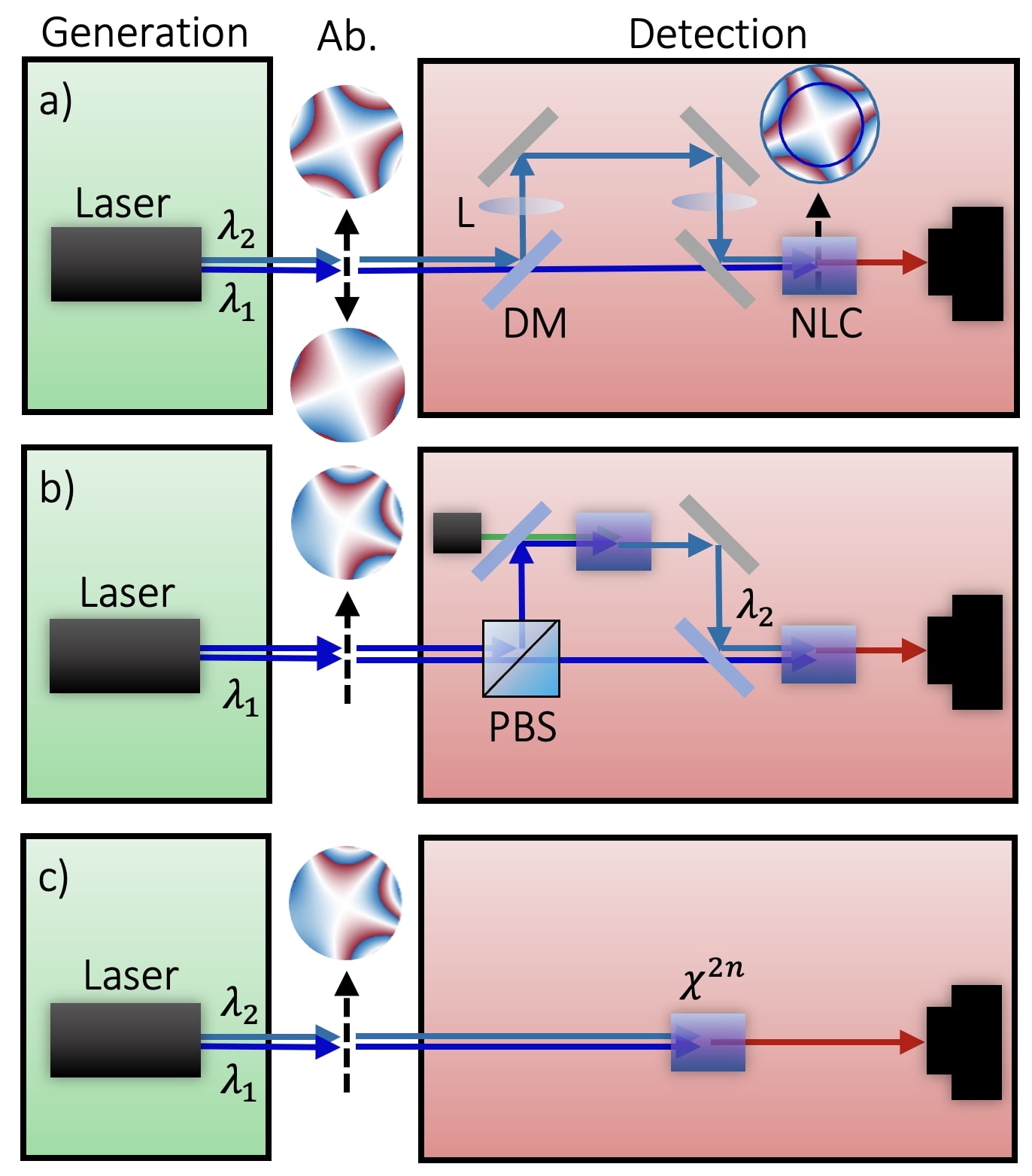}
\caption{Using nonlinear optics so that light corrects light is versatile in implementation where (a) resizing the the input beams of different wavelengths corrects the primary aberration, (b) using a second nonlinear process allows the same aberrations to be incurred by the signal and probe beams and (c) the principle of phase conjugation can be extended to any crystal with an even-ordered non-linear susceptibility. DM, dichroic mirror; L, lens; NLC, nonlinear crystal; PBS, polarising beamsplitter.}
\label{Fig6}
\end{figure}
%%%%%%%%%%%%%%%%%%%%%%%%%%%%%%%%%%%%%%%%%%%%%%%%%%%%%%%%%%%%%%%%%%%%%%%%%%%%%%%%%%%%%%%%%%%%%%%%%%%%%%%%%%%%%%%

Interestingly, we show that after undergoing distortion, resizing one beam relative to the other achieves a change in the respective aberrational strength overlapping with the unsized beam. To do so, a mismatch of $\beta = \frac{w_2}{w_1} = 1.4663$ between the probe ($w_2$) and signal ($w_1$) beam waists were made when demagnified onto the crystal in Fig. \ref{Fig1}. Using the similarity ($S = \frac{\left[\sum_{x,y}{\sqrt{I_{NA}(x,y)I_{Cor}(x,y)}}\right]^2}{\sum_{x,y}I_{NA}(x,y)\sum_{x,y}I_{Cor}(x,y)}$) between the measured unaberrated ($I_{NA}(x,y)$) and probe-corrected ($I_{Cor}(x,y)$) DFG intensity distributions, we quantify the correction efficacy for a range of aberration strengths encoded on the probe, while the signal remained fixed. $S = 1$ ($S <1$) indicates perfect correction (presence of uncorrected aberrations). For generality, three cases where the aberrational Zn order, magnitude of the aberration on the signal and spatial mode were tested and shown in Fig.~\ref{Fig5}. In each case, we find the same aberrational strengths do not cancel the distortion in the generated beam. More specifically, in Fig.~\ref{Fig5} (a), we find an astigmatic LG$_{\ell = 1}$ signal beam with a coefficient (strength) of $C_2 = 10$ requires a coefficient of $C_1 = 20.8$ on the probe to cancel the distortion due to weakening of the relative strength from the beam enlargement. For qualitative comparison, insets (i), (ii) and (iii) give the unaberrated DFG beam, aberrated DFG beam and aberrating mode, while the optimally corrected DFG beam is shown as the inset in the plot. Next, both the aberration type (trefoil) and strength ($C_2 = 15$) were altered in Fig.~\ref{Fig5} (b), giving optimal correction with $C_1 = 43.9$ and in Fig.~\ref{Fig5} (c), an astigmatic LG$_{\ell = 2}$ with the same strength as (a) needs approximately the same strength ($C_1 = 21.4$). 

Here the Zernike radial orders (e.g. astigmatism with order 2 and trefoil with order 3) do not scale linearly with $r$, but instead to the power of the order ($m$) from exponentiation of the radial term in the function. One can then derive the relation, $\frac{C_2}{C_1} = \beta^{m}$ dictating the relative strength change as the result of resizing the probe. From this, we find $\beta = $1.443, 1.431 and 1.466 for each case in Fig. \ref{Fig5} agrees well with the experimental demagnification ratio (1.466). This confirms that resizing the input modes relative to each other allows one to employ a corrective strength to compensate for the variation in wavelength traversing a distorting medium. \BS{Intuitively, such resizing alters the phase gradient seen by the other beam and thus can be used to perfectly correct for the primary (major) aberration in any system of interest, as illustrated in Fig. \ref{Fig6} (a). As an example, phase insets (Ab.) show astigmatism aberrations incurred by wavelengths that have a factor of $\alpha = 2$ difference between them. By splitting, resizing and recombining the aberrated beams correctly, a perfectly corrective overlap is formed in the crystal (insets above NLC show the resized beam phases which match perfectly) as part of the nonlinear detection system. One may be cognisant that in many aberration correction strategies, one can also find limitations in compensating for all the aberrations present and thus achieving correction of the aberrations is limited to the prominent contributors \cite{hampson2021adaptive}. Such corrections, however, still provides a significant improvement in the systems, facilitating satisfactory performance. Accordingly, employing our resizing approach to facilitate measurement-free correction of the prominent aberration indicates a promising approach to optical systems using spatial modes that undergo distortions such as in communications or metrology. One my additionally utilise the wide ranges of nonlinear crystals \cite{nikogosyan2006nonlinear,luo2019recent,tran2016deep} to compensate for the disparity by making the input wavelengths closer, but at the expense of detection at much higher (mid- or infrared) wavelengths.}

\BS{Alternatively, additional strategies can be engineered to improve the efficacy of the measurement-free approach, each implementing the core of our approach. Here, one can move from resizing the beams in the detection box to a two-step nonlinear system such that the probe is the same wavelength as the signal and thus incur identical distortions (Fig. \ref{Fig6} (b)). On the detection side, after separation with a property such as polarisation, the aberrated probe can undergo an initial nonlinear process to create a seondary probe at the desired wavelength for DFG (retaining the aberration by using another unstructured pump). The new probe is then recombined with the signal using a dichroic mirror for the DFG process in a second crystal. Furthermore, as illustrated in Fig. \ref{Fig6} (c), one need not be restricted to utilising phase conjugation from second-order nonlinear processes ($\chi^{2}$). It can be shown (see Appendix) that our core principle principle of phase cancellation is true for parametric wave mixing processes of even order ($\chi^{2n}$) in higher-order difference wave mixing, keeping in mind that the encoded phases of interest ($\Phi_\text{info}$) are generated with a related factor ($\Phi_{G} = 2n\Phi_\text{info}$). Furthermore, one may add the phase-conjugating crystals in a cascaded configuration (see Appendix), making it possible to further extend our scheme to wavelength combinations and processes not possible in the straightforward approach.}

\begin{figure}[ht!]
\centering\includegraphics[width=8cm]{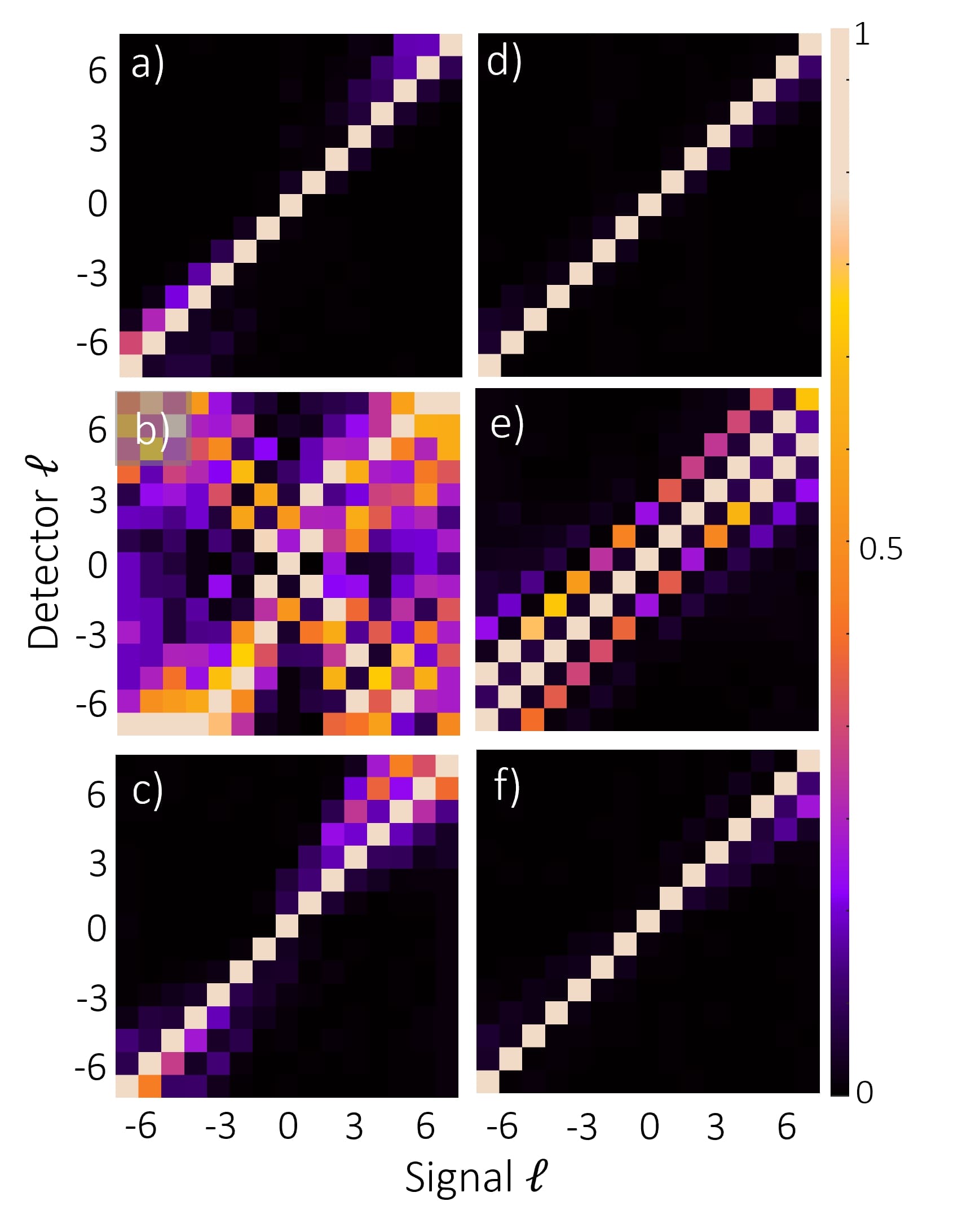}
\caption{The probe field is used as a detector for OAM modes of $\ell = [-7,7]$ in the cases where the beam size expands as dictated by the OAM value (a-c) and a size-adjustment of $\frac{w}{\sqrt{(\ell+1)}}$ included to mitigate the OAM-dependent expansion in the generated modes. Detection cross-talk matricies of the system are shown (a,d) without applied aberrations, (b,e) with the 4-mode Zernike aberration and (c,f) with the aberrations corrected. Each row is normalised with the maximum value.}
\end{figure}
Lastly, we demonstrate a prepare-and-measure system that allows us to retrieve the correct encoded modes despite the presence of distortions. Here the conjugating nature allows not only the phase distortions to be eliminated, but the phase of equivalent spatial modes as well. For instance, only when the same OAM mode is encoded on the signal and probe does the DFG beam contains a flat phase. This forms a Gaussian intensity distribution in the far-field which results in the presence of an on-axis intensity. Such matching of input modes (from the orthogonality relation of LG modes) allows it to also be used as a spatial mode detector. For continuity, we chose the same previous four-state Zernike superposition to be the aberration and show how such a detection system without aberration, with an aberrated signal beam and with a corrected detector mode (probe) performs. We do so in the case where the OAM beam modal profiles expands naturally with $\ell$ and with a mitigation of this expansion by encoding a size-adjustment of $\frac{w}{\sqrt{(\ell+1)}}$ the each OAM mode. In the first case, observation of the detection system for the ground-truth (before aberrations) may be noted as having some higher mode cross-talk in the detection matrix (a), but largely detects the correct encoded OAM. However, with aberrational effects added to the signal, one is not able to distinguish the modes sent as seen with cross-talk extending to adjacent modes and forming a cross-diagonal pattern. Applied corrections on the detection beam retrieves the detection diagonal, although it begins to degrade as the higher-order modes are used. This can be attributed to the enlarged sizes on both the detection and signal beams causing additional aberrations and mismatch being accumulated throughout the optical system for both the beams. This detracts from the encoded and corrected aberrations. Confirmation of this may be observed in the case where the expansion of the beams were mitigated in (d-f) where the ground-truth detection matrix (d) already demonstrates an improvement in the system. The aberrational effects in (e) are additionally mitigated, but a clear distortion of the information being sent is still present, where adjacent modes are detected along with the modes being sent. Application of the correction on the detector mode then fixes the aberrational effects to yield the detection of the correct modes (f), in close agreement to what was observed for the non-aberrational case in (d).

\section{Conclusion}

In conclusion, we demonstrated the ability to use light as a mechanism to correct aberrated light modes through difference frequency generation. An advantage of the DFG mixing process is the correction need only be in the form of identical aberrations, providing, in principle, the opportunity to naturally cancel any incurred distortions when both the input beams (one carrying a desired structure and the other used for conversion of the structure) experience the same aberration. We demonstrated this principle for a wide array of both spatial modes and aberrations, starting from azimuthally-varying aberrations to including radial variations and linear combinations thereof. In doing so, excellent restoration of the spatial states were found for Laguerre-Gaussian modes with a range of OAM and a symmetric superpositions thereof, forming exemplary complex structures that hold utility across a wide range of applications. We showed that this technique can be important to other higher order nonlinear phenomena or cascaded nonlinear effects, enabling improved wavelength manipulation.

Furthermore, the practical aspects of different wavelengths traversing the same aberrating medium were considered, in the event measurement-free correction is desired, and found a relative resizing of the inputs can mitigate the disparity in the strength of the aberrations incurred. In doing so, while not all aberrations can be corrected, the primary contribution can be cancelled in the DFG mode according to the order one wishes to compensate for. We also consider the ability to use similar wavelengths in the process, where the wavelength-dependent disparity can be reduced at the cost of needing to detect DFG light in the IR region. Finally, by employing an identically aberrated detector beam, we were able to restore the ability to detect the encoded modes. Here, good agreement was found between the detected and encoded modes that were left to scale in size with the OAM charge. Compensating for the scaling in the modal set then further improved the restoration of the encoded states. Application in such a system would be then be useful for retrieving information through noisy channels. Notably, the probing mechanism reliance on light itself renders it advantageous under rapidly varying distortions, such as atmospheric turbulence, making this technique a valuable tool for various applications in optical communications to imaging and sensing. Furthermore, for a non-degenerate wavelength setup as used here, one is afforded the ability to detect in the visible range when working with information carried by structured light in the difficult-to-detect near infrared region. 

\begin{backmatter}
\bmsection{Funding}
The authors acknowledge the funding from the Department of Science and Innovation (DSI) as well as the National Research Foundation (NRF) in South Africa. Support from the Italian Ministry of Research (MUR) through the PRIN 2017 project “Interacting photons in polariton circuits” (INPhoPOL) and the PNRR project PE0000023-NQSTI is acknowledged. We also acknowledge support from the Italian Space Agency (ASI) through the "High dimensional quantum information" (HDQI) project.

\bmsection{Acknowledgments}
The authors would like to thank Dr Isaac Nape and Dr Paola Concha Obando for useful discussions.

\bmsection{Disclosures}
\noindent The authors declare no conflicts of interest.

\bmsection{Data availability} Data underlying the results presented in this paper may be obtained from the authors upon reasonable request.

% \bmsection{Supplemental document}
% See Supplement 1 for supporting content. 

\end{backmatter}

%%%%%%%%%% If using BibTeX:
\bibliography{Biblio}

\section*{Appendix}
\WB{
We can expand the concept and demonstrate that this correction is not exclusive to second order nonlinear processes, but a general feature of parametric wave mixing processes of even orders. In a general form, the nonlinear response of a medium to an incident field is of the form
\begin{equation}
	\bm{P}^{(n+m)} = \chi^{(n+m)} \left(\prod_{i=1}^{n}\bm{E}\right) \left( \prod_{j=n+1}^{n+m}\bm{E}^{*}\right) 
\end{equation}
where $\bm{P}^{(n+m)}$ is the nonlinear polarization of the medium and  $\chi^{(n+m)}$ is the susceptibility tensor \cite{buono2020chiral}. The first and second products represent upconversion and downconversion processes respectively where the appropriate tensor components should be considered. Without loss of generality, any scalar component of an even nonlinear process involving input fields $\bm{E}_1$ and $\bm{E}_2$ can be written as
\begin{equation}
\bm{P}^{(2n)}=\chi^{(2n)}\left(E_1+E_2\right)^{n}\left(E_1^*+E_2^*\right)^{n} \ .
\end{equation}
Noticeably, there will be terms responsible for the excitation of fields with multiples of $\omega_{\text{G}}=\omega_1-\omega_2$ which are the terms proportional to the powers of the product $E_1E_2^*$. These are referred to as Higher Order Difference Wave Mixing (HODWM)\cite{gaarde1996theory} and represent the absorption of multiple photons at once.
We can write them as
\begin{equation}
	\bm{P}^{(2n)}(n \omega_{\text{G}})=\chi^{(2n)}M_1^n M_2^{n*}=A_1^nA_2^ne^{in(\Phi_1-\Phi_2)} \ .
\end{equation}
}
\WB{
Similarly to the initial argument, if $\Phi_1=\Phi_\text{info}+\Phi_\text{Ab}$ and $\Phi_2=\Phi_\text{Ab}$ then the generated phase profile would be $\Phi_\text{G}=n\Phi_\text{info}$. Additionally, by setting the absolute part $A_2$ to be spatially uniform, we see that the generated field $M_\text{G}^{(2n)}=M_1^n$, meaning that in this case the generated field is a positive integer power of the signal field. For the case of spatial transverse modes such as LG and Hermite-Gaussian beams, these fields can be seen as multiple products of modes, for which there are one-to-one correspondences \cite{kotlyar2022product,wu2022conformal,alves2018conditions} that essentially map the resulting field back to the original message unambiguously. In addition, this nonlinear dependence has been shown to be advantageous in detection processes \cite{qiu2018spiral}.
}

\WB{
As long as diffraction effects are negligible, this is true not only for a single process of DFG but also for cascaded nonlinear processes: every $N_h$-th harmonic generation of $E_1$ would add an integer $N_h$ multiple of the aberration profile to the phase profile - the same being true for $E_2$ in its $N_k$-th harmonic. A number $N_c$ of cascade of DFG processes combines harmonics of $E_1$ and $E_2$, thus applying a partial aberration correction $N_c$ times. The total aberration correction is achieved when $N_h=(N_c-1)+N_k$. 
It is important to notice that the combination of harmonics generation and DFG processes can be achieved in different media: instead of a single crystal or gas it is possible to use a sequence of crystals or gas chambers combined by imaging systems. This can be used as method to optimize the efficiency of a certain process or enable specific wavelengths which are not possible with a single process. This includes the case where a corrected beam is generated in the same wavelength as the original signal after two or more nonlinear interactions. In this case, a total depletion regime can be used for a complete substitution of the aberrated signal for a corrected beam.
}

\end{document}